# A Matched Pairs Analysis of International Protection Outcomes in Ireland

Gerard Keogh

Central Statistics Office,

Dublin 6, Ireland

*email: gerard.keogh@cso.ie

Abstract

We examine over 40,000 International Protection (IP) determinations for non-EEA nationals covering a 16 year period in Ireland. We reconfigure these individual outcomes into a set of over 23,000 matched pairs based on combination of direct matching and propensity score matching. A key feature of this approach is that it replicates the statistical features of an experimental set-up where observational data only are to hand. As a consequence we are able to identify those explanatory factors that in fact contribute to the grant of IP. This is a key innovation in the analysis of protection outcomes.

We centre our study in the realm of International Relations studies on protection. We are particularly interested in whether immigration policy is a latent tool used to influence the odds of a grant of IP, specifically via the introduction of the Immigration Act 2004. Using both conditional maximum likelihood and mixed effects models we find this is not the case; this conclusion is both novel and profound in a matched pair context. On this basis we conclude there can be little justification for the perception that immigration policy is a latent tool affecting the protection process in Ireland.

Keywords: International Relations, Geneva Convention, Refugee, International Protection Outcomes, Matched pairs, Conditional Maximum Likelihood Logistic Model, Mixed Effects Logistic Model.

## 1. Introduction and Philosophical Perspectives on Asylum Determinations

In the EU two complementary forms of protection are available to those seeking asylum, the Geneva Convention (GC) and Subsidiary Protection (SP). Collectively these two forms of protection are known as International Protection (IP). The GC and its optional 1967 Protocol applies to persons in need of protection who are considered part of a group based on nationality, race, religion, or political affiliation. Meanwhile, SP also covers individuals seeking asylum on their own account, e.g. a person at a real risk of suffering serious harm if they remain in their home country. The GC, which was proclaimed by the UN General Assembly in 1951 in the aftermath of the horror of World War II, sets out basic rights and obligations of states in regard to asylum seekers and refugees. Chief among these rights is the norm of *non-refoulment,* this prohibits states from forcibly returning individuals who fear a return to their country of origin. As of 2015 one hundred and forty eight states are party to the GC. In the EU approximately 1.1 million persons are currently afforded protection as recognised GC refugees, while Ireland hosts over 6,000 GC refugees with many thousands already having been granted citizenship.

Unsurprisingly, the humanitarian idealism of fostering the rights of asylum seekers and refugees that is reflected in the spirit of the GC, sometimes conflicts with the day-to-day reality of nation states self-interest. In the realm of International Relations (IR) studies, a more formal philosophical framework known as the constructivist vs. realist debate exists for the discourse surrounding this conflict. The constructivist viewpoint [17] sees human development as a gradual inexorable rational process that emphasises the emergence and diffusion of ideas, which eventually become the norms, laws and institutions of a united and good civil society. In contrast to the philosophical idealism underpinning constructivism, others see the development of society in more limited terms based on the notion of self-interested cooperation of individuals. This viewpoint is known as (neo)-realism [27]. Basically, this is as a minimum code of law that restrains people from harming one another but otherwise leaves people free to do whatever the law does not expressly forbid. In relation to asylum and refugee concerns this tug-of-war between 'norms' versus 'interests' began from the moment the GC was proclaimed in 1951 and continues with fervour to this day [21]. Accordingly, there is a perceived gap between commitment to IP on the on hand and narrow compliance with protection treaties on the other. Naturally, this tug-of-war between compliance and commitment has resulted in concerns being raised about the outcome of protection cases in general. This leads some who advocate for asylum seekers and refugees to question the reliability of existing arrangements and procedures to determine claims for protection. Evidence to support this view is provided in [2] and [13]. They report that determinations in the US may depend on arbitrary factors such as the geographical location of the deciding judge, or that an outcome may depend largely on chance (e.g. on the deciding judge assigned), or the outcome may even depend on foreign policy concerns. Indeed, a leading study [22] of refugee application outcomes in the US describes this phenomenon as 'refugee roulette'.

However, in Ireland and in many EU countries this controversy seems less hot as there is a substantial belief that the process of European integration, that is the development of European institutions, has facilitated domestic asylum policy change [25]. Yet, even in this environment, Non-Governmental Organisations (NGOs) advocating for asylum seekers from a constructivist viewpoint, still seem to retain the perception that asylum policy may be driven by latent factors beyond those relevant to a particular asylum seeker's case. Evidently, action by officials involved in Justice and Home Affairs, who in this context may be termed realist state actors, has contributed to this view. Indeed, the operation of the Common Travel Area between Ireland and the UK, the commencement of the Schengen Borders Code (Regulation 562/2006) and the EU Dublin II Directive 2003/343, which may be thought of as latent asylum system management mechanisms, have fostered the notion of "fortress Europe".

However, the notion of "fortress Europe" misses the key underlying point faced by realist state-actors operating in the protection sphere. The policy issue at stake is the pressing issue of "burden sharing", which has been central to the 2015 Syrian refugee crisis. Thus, the tug-of-war between constructivist actors and realist state actors comes down to constructivists advocating for common and better asylum procedures and greater access at EU frontiers. Meanwhile, realist state actors pursue the goal of common asylum procedures, but may implicitly use border and immigration control procedures as a device to dodge the thorny topic of burden sharing. Interestingly, the common ground between the viewpoints is in the area of asylum procedures. When this is taken in isolation from wider concerns, it provides an explanation for the "absence of restrictiveness" in relation to EU asylum policy, as observed in [20]. While realists may take comfort from this narrow view, a more thorough analysis should take account of latent influences that may affect domestic actors on the ground making protection determinations. The focus of this paper is to see whether latent policy efforts affect outcomes for asylum seekers. We quantitatively examine outcomes of individual asylum cases to see whether and to what extent those latent factors may have impacted on Ireland's openness to asylum seekers. Accordingly, we also seek to gauge the validity of NGO perceptions, with a view to discovering whether or not a narrow interpretation of protection law may have existed in Ireland.

In recent years there have been a number of quantitative studies of individual case level protection outcomes and their associated policy implications, see [2], [3], [13], [14], [19], [22]. Broadly these studies tend to agree that asylum determination systems are strict, but there is less support for the view that systems may be biased against the asylum seeker. By biased we mean factors largely under the control of Justice and Home Affairs actors alone, such as whether the applicant is given an interview, and not asylum applicant specific factors like gender, substantially determine the case outcome. Of course the studies listed above all provide observational analysis of the data and consequently the validity of their conclusions can only be suggestive. This fact further weakens any support for the claim that asylum determination systems may or may not be biased. Accordingly, to validate this claim a stronger level of proof is desirable. In this paper we revisit the Irish dataset used in [19] and reconfigure it as a set of matched pair case-controls. This observations dataset comprises over 40,000 individual protection outcomes made in Ireland in the period 1998 to 2013 across 159 (non-EEA) nationalities. From this observational dataset we construct matched pairs using mixed approach. This involves matching on specific factors such as nationality and gender, etc., as well as using propensity score matching for other covariates [10]. We then model the odds of a grant of protection against a vector of independent normative factors describing the policy environment, in order to see if any of these provide statistically significant evidence that changes the odds of a positive (grant) outcome.

As in [19] our main hypothesis of interest is whether and to what extent policy instruments may impact on Ireland's generosity to asylum seekers as recorded in outcomes in individual asylum cases. Clearly, if true, then NGO perceptions mentioned above have validity and a narrow interpretation of protection law possibly based on self-interest exists. However, if false, then perceptions are also false and Ireland is a country 'open' to asylum seekers where policy makers apply protection law fairly. Accordingly, this study quantitatively explores this tug-of-war between the idealism of an open welcome for asylum seekers and refugees, and the reality that may be practiced by those making protection determinations. We situate our analysis within the constructivist vs. realist debate in IP with a view to more comprehensive qualitative exploration. We stress that this analysis contributes in a novel way to this debate, since any factor that is statistically significant within our matched pairs case-control analysis, may be thought of as providing evidence in favour of a contributory relationship between the factor and a positive IP outcome. Accordingly, this study extends the analysis reported in [19] in a way that provides a meaningfully stronger level of proof about the performance of Ireland's asylum determination system and so enables us to consider whether, as a consequence, Ireland tends to view asylum through the lens of burden sharing.

The remainder of our paper is organised as follows. In section 2 we highlight some key features of the IP framework in Ireland and outline the philosophical perspectives relevant to the IP debate. In section 3 we describe how we construct the matched pair case-controls from the observational data. Section 4 sets out the statistical models we use to analyse the matched pairs. In section 5 we analyse and discuss the results obtained from fitting the models described in section 4 to the matched pairs data. Section 6 concludes.

## 2. A summary of the framework and (latent) policy issues in International Protection

**The International Protection framework**

The international framework for asylum policy is the 1951 Geneva Convention (GC) and its optional 1967 protocol. According to the GC "a refugee is a person who owing to a well-founded fear of being persecuted for reason of race, religion, nationality, membership of a social or political group, is outside the country of his nationality, and is unable to, owing to such fear, or is unwilling to avail himself of the protection of that country". As noted in the Introduction such persons enjoy the right of *non-refoulment* – that is they cannot be expelled or returned in any manner whatsoever, to the frontiers of territories where their life or freedom is threatened. A person who is granted refugee status under the Convention is guaranteed protection by that state and is generally referred to as a Geneva Convention Refugee or simply as a Convention Refugee.

In the EU the framework underpinning asylum policy is the Common European Asylum System (CEAS). This is based on three pillars; first, bringing more harmonisation to standards of protection by further aligning EU States' asylum legislation; second, ensuring effective and well-supported practical cooperation; and third, increasing solidarity and responsibility among EU States. The CEAS has four important legislative aspects:

- the Directive on Reception Conditions for Asylum Seekers 2013/33/EU
- the Qualification Directive 2004/83/EC
- the Asylum Procedures Directive 2013/32/EU
- the Dublin II Regulation 2003/EC/343.

Further details on all aspects of the CEAS and its development since 2005 are given in [9] while more extensive summaries are also given in [18] and [19]. Many EU states also provide a non-harmonised protection status, this is usually called humanitarian leave to remain and does not fall directly within the realm of the CEAS.

Unlike most EU states, Ireland in the years prior to 2016 did not operate a so-called 'single procedure' for determining protection claims. Basically, a single procedure considers all forms of protection concurrently. In Ireland, a claim for IP is first considered under the GC which is transposed into national law in the Refugee Act (1996). As of writing this article, the Refugee Act is being superseded the International Protection Act (2015) which provides for a single procedure. In Ireland during the period 1998 to 2013, an asylum applicant progresses sequentially through stages beginning with a first instance application to the Refugee Applications Commissioner (RAC), followed in the event of refusal by a right to appeal to the Refugee Appeals Tribunal. If the application for GC refugee status fails the applicant may then apply for Subsidiary Protection (SP). The SP process was commenced in Ireland in 2006 under the Immigration Act 2004, however and prior to 2014 where an SP application failed there was no right of appeal. An applicant who fails the SP test is finally invited to make a submission to the Minister for Justice and Equality, for permission to stay in Ireland on humanitarian grounds; this non-harmonised humanitarian process is outside the realm of IP.

**Policy issues with a potential latent influence on International Protection decisions**

Certain policy efforts can bear heavily on protection decisions. The focus of this paper is to see whether latent policy efforts affect outcomes for asylum seekers. Latent policies are mechanisms that indirectly manage or influence protection determinations. Broadly speaking latent policies are actions in the immigration sphere that limit access to a country or they may be efforts that could implicitly reinforce preconceptions. Clearly, any evidence of latent policies is relevant in the tug-of-war between commitment and compliance in the protection sphere. In Ireland's case the main latent policy mechanism that may influence an IP outcome is the Immigration Act 2004.

Ideally, an immigration policy instrument should ensure those with little or no basis for an asylum claim will find it hard to gain entry. If this is so, then we should see a greater proportion of asylum claims resulting in a grant of refugee status. As a consequence, the overall odds of a positive outcome (i.e. grant of IP) should increase after an immigration policy instrument is introduced. Interestingly, we suggest this is the explanation for the likelihood of a grant actually increasing following the introduction of IIRIRA (Illegal Immigration Reform and Immigration Responsibility Act 1996) and Real ID Act in the US, as reported in [2]. In Ireland the Immigration Act 2004 is the key piece of legislation aimed at managing access and border control. Thus, if the recognition rate, i.e. the proportion of grants to all asylum determinations in a year, increased after 2004, then officials will view their actions as both successful and humane. It will also demonstrate to NGOs that state actors are behaving in a manner that shows their commitment to human rights principles, while having due regard for the rights of their own citizens. However, if the recognition rate is observed to fall, then the actions realist state actors will have unfairly hit bone fide asylum seekers. In this scenario, NGO perceptions that Ireland only complies with a narrow interpretation of the GC and is not committed to it in spirit will be justified. In this study we explore whether this has occurred by seeing if the odds of a grant of IP has changed in response to factors associated with the implementation of the Immigration Act 2004.

Evidently, each determination to grant or refuse protection is based on a wide set of criteria. It includes the applicant's testimony, their medical condition and frame of mind, and applicant specific criteria such as age and gender etc. This 'case file' of information, most of which unfortunately is not accessible to us, is defined by the UNHCR [26] as the credibility assessment of the asylum claim. It is clear that a decision to grant or refuse IP therefore depends critically on the overall effect of the information in case file on the deciding officer or judge. Thus, while each individual factor associated with the immigration act may be relevant on its own, it also seems reasonable to assert that a group of immigration related factors may act together on the outcome. Clearly, the overall value of this information can shed light on the commitment and compliance debate, and interestingly we can reflect its importance in determining the outcome. Thus, in a matched case-control situation, if the overall effect of factors associated with the immigration act is substantial, relative to overall effect of applicant specific factors, then immigration is likely being used to limit access to asylum. If this is the case, then the perceptions of NGOs and others, who may take the view that officials are using latent methods to influence asylum outcomes, may to some extent be justified.

Of course, conditions in the protection applicant's country of origin are crucial in any IP determination. Here, the independent country of origin information (see www.ecoi.net) and the asylum seeker's account of their experience in their home country inform the credibility of an asylum seekers claim for protection. As a consequence, the applicant's stated nationality is central to determining the IP outcome - previous studies [2], [19] and [22] have also shown this turns out to be the case in practice. Nonetheless, latent immigration policies may implicitly reinforce perceptions about certain nationalities and in so doing adversely contribute to biasing the odds of a grant of IP. Clearly, if this is evident in the odds of a grant of IP, then it will raise doubts about the even-handedness of officials involved in making protection decisions and their commitment to fair and humane procedures. Intriguingly, insight into this topic is available by comparing population average odds of a grant of IP to average odds when we adjust for nationality. Given that nationality is central to determining the IP outcome, we therefore should see a substantial change in odds ratios when we adjust for the size of each nationality. If this fails happen then IP outcomes for certain nationalities are not determined by nationality in a smooth manner, equivalently, outcome odds are non-smooth functions of nationality specific sufficient statistics. This of course means that some hidden effect that remains unknown to us is confounding the outcome for certain nationalities, and this hidden effect may be latent immigration policy. We explore this by examining whether there is evidence to support the view that the Immigration Act 2004 has disproportionally influenced the odds of a grant of IP, for nationality adjusted matched pairs compared to population level average odds.

## 3. Construction of matching pair cases

First we must construct the set of matched cases from the observations dataset of 40,437 individual protection decisions. On this observations dataset each individual IP final outcome/determination for each applicant, is recorded as a Grant=1 (whether it be a GC outcome or an SP outcome) or final Refuse=0 (at GC where it applies or at SP stage). Withdrawn cases, those who apply for protection and either withdraw voluntarily or fail to cooperate with the determination procedure and are 'deemed withdrawn', all result in a negative IP outcome, these cases are excluded. We construct our set of matched pairs from the set of observations using a mixture of matching and propensity scores (see [10] Chapter 9).

Our matching variables are those variables that must be the same for the case and the control; the list of variables on the observations dataset is shown in Table 1. In our case the matching variables are 'Nationality', 'Gender', 'Adult' and 'After 2004'. The matching variable 'After 2004' is included because it distinguishes whether the application occurred before or after the commencement of the Immigration Act 2004. For the remaining variables on the dataset we look for matches using the propensity score method [10]. Within each 'Nationality' by 'Gender' by 'Adult' by 'After 2004' group, we find the 'best' one-way linear model based on the deviance statistics [Dob], by regressing the 'Outcome' sequentially on all other variables listed in Table 1. In the manner of [19], we include 'Nationality' as a one-way random effects variable to reflect the importance of nationality in the determination procedure. PROC GLIMMIX is SAS is used to fit each of these sequential models. Figure 1 shows a plot of the deviance that results as each variable is added. Based on this plot we see the variables 'Year (of determination)', 'SP decision', 'RIA', 'Unaccompanied Minor', 'Interviewed', 'Ever Married' and 'GDP Ratio' contribute substantively to increasing the deviance. We compute propensity scores using this set of one-way fixed effect variables and 'Nationality' as a one-way random effect.

We proceed to construct our matched cases by selecting each observation that resulted in a grant (i.e. Outcome=1) in turn. For each of these we select the set of observations that resulted in a refusal (i.e. Outcome = 0) and find a matching group of refused observations based on 'Nationality', 'Gender', 'Adult' and 'After 2004'. Within this refused group, we then select a specific matching control based on the closest matching propensity score, between the selected case and each observation in the refused group of observations. When a match is selected from the refusal set is it excluded from subsequent selection. This assignment matching mechanism generates a set of 1-to-1 matches. However, 1:1 assignment only uses about 25% of the available refuse observations, so there is a significant loss of information as about three-quarters of the refuse observations are ignored. To ameliorate this we in fact take the top three propensity score matches thereby generating a set of 1:3 matched pairs, this means we actually use about three-quarters of the refuse observations, thereby greatly limiting the information loss.

Table 1
Variable Description

| Function | Name | Description |
| --- | --- | --- |
| Outcome | IP Determination Outcome | Final Outcome of IP application |
| Matching | Nationality | Random Effects variable to allow outcome to vary randomly depending on nationality |
| | Gender (female) | Female = 0; Male = 1 |
| | Adult | Under 18 = 0; 18 or over = 1 |
| | After 2004 | Immigration Act 2004 – a factor variable to control for cases before vs. after the introduction of this Act |
| Process specific | SP Decision (Yes) | Subsidiary Protecton (SP) - whether the final outcome was based on an SP application |
| | Unaccompanied Minor | If the person was an unaccompanied minor - these account for about 3.5% of Ireland's determinations |
| | Interviewed | If the applicant underwent an interview |
| | Free Country of Origin | If the applicant's country of origin is designated by Freedom House as 'Free or partially free' |
| | At Risk (political terror scale) | If the applicant's country of origin is designated as 'at-risk' based on a rating of 4 or 5 on Gibney's Political Terror Scale |
| | Refused Leave to Land | Number refused leave to land in the determination year by nationality |
| | Returned to Country of Origin | Number of persons deported in the determination year by nationality |
| Applicant specific | Age (Years) | Age in years at determination |
| | RIA (Yes) | Whether the applicant spent some time in a Reception and Integration Agency (RIA) centre |
| | Air Travel (Yes) | If they travelled by air |
| | Asylum Reason (political) | If applicant stated their asylum application was based on political grounds |
| | Religion (stated) | If they stated a religion |
| | Ethnicity (stated) | If they gave an ethnicity |
| | Ever Married (Yes) | Iif the applicant stated they had been married at some point |
| | English Speaking (Yes) | If the applicant could speak English |
| Other | Year (of determination) | The year in which the determination was made |
| | Length (3Years) | The length of time between application and final determination less than 3 years= 0; 1 otherwise |
| | GDP Ratio | Ratio of GDP in the home country to GDP in Ireland in the year of application Upper Quartile = 1; 0 otherwise |

Two features of this matching mechanism are worth highlighting. First, propensity score matching ensures the matching assignment mechanism is ignorable [10]. Broadly speaking, this means the distribution of matched case-controls from observational data, is the same as the distribution that would have resulted had the case-control assignment been completely randomised. In short, propensity score matching recreates the conditions of a completely randomised experiment, which of course is a prerequisite for the level of contributory proof we seek. Second, we note by design this matching mechanism produces a case-control dataset where the case outcome is a grant and the control outcome is a refusal. We are thus interested in measuring the effect of the treatment, i.e. the change in conditions or factors influencing the determination, that result in a positive grant outcome, given initial conditions or factors that would have resulted in a refusal. The case-control dataset that results from this matching process has 24,156 matched pairs covering 96 of the original 159 nationalities.

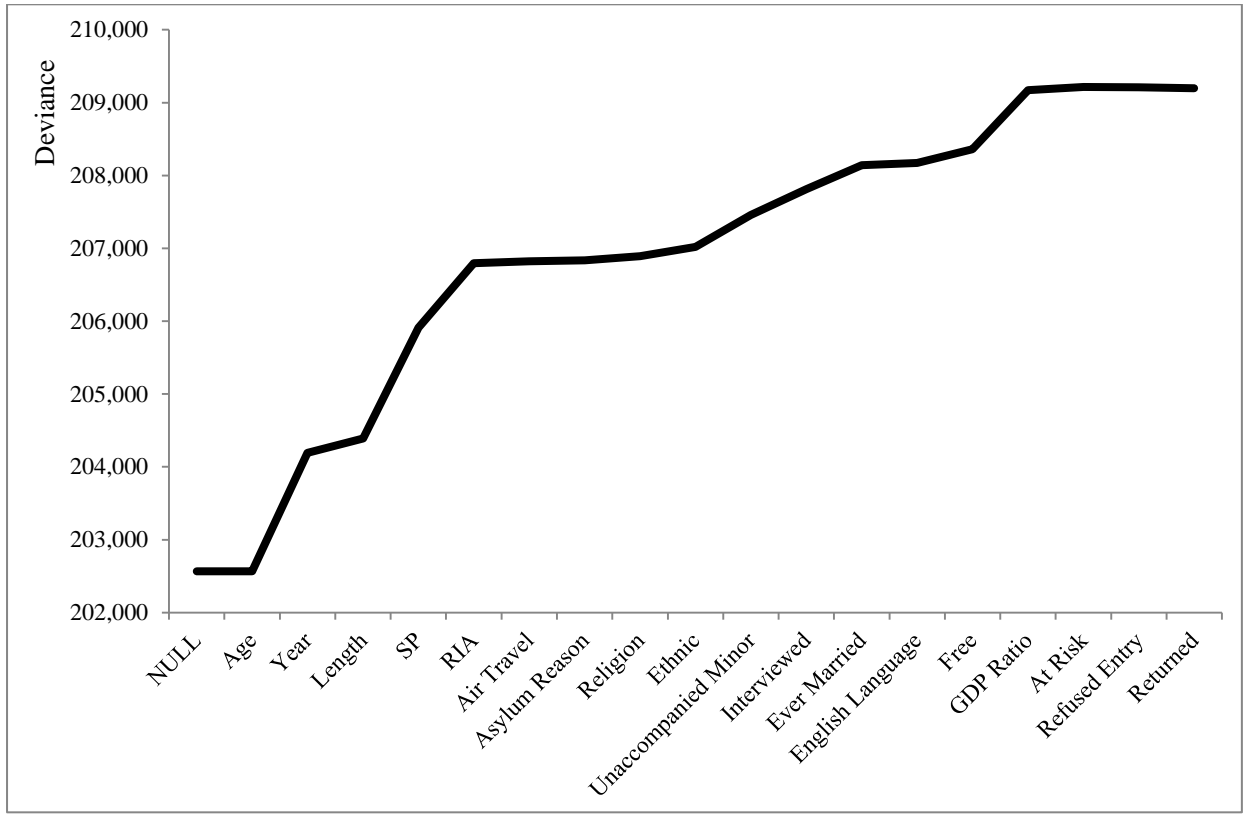

Fig. 1. Deviance plot from the regression of Outcome on explanatory factor

It is also worth noting that this this set of matched pairs is constructed on the basis that the variables are not subject to any logical constraints. Of course, in reality this would not be the situation for matched case-controls, where the outcome is the difference between a before and after measurement on the same person or sampling unit. Thus, for example, in practice a person who is not an unaccompanied minor as a control (i.e. before) cannot be a matched unaccompanied minor as a case (after). Equally, a person who is given an interview as a case must also be given an interview as a matched control, or indeed, the age of the control must be less than or equal to that of the case. Interestingly, if we impose stringent before-after type variable constraints similar to those mentioned on all variables, we get a dataset of fewer than 200 matching cases controls. Unfortunately, this dataset has poor coverage across key variables such as nationality. Imposing variable constraints therefore renders the 24,156 matched pairs dataset virtually useless for further analysis. Accordingly, other than for imposing the unaccompanied minor constraint, we revert to imposing no variable constraints on our matched dataset, and instead treat our dataset as if they were simply a set of pairs. Specifically, we have in mind the matched cases are like a set of 'husband and wife' style pairs, with the condition that the 'wife' is the control having a refusal (Outcome=0) while the 'husband' is the case having a grant (Outcome=1), or vice versa. Clearly, under this design the difference between the case and control outcomes is 1 for all 23,246 matched pairs satisfying the unaccompanied minor constraint. Our case-control analysis proceeds with this paired set of 23,246 case-controls which still covers 96 nationalities across all years from 1998 to 2013.

## 4. Modelling Methods

Interestingly, data from a case-control study can be analysed using a logistic regression model to measure the effect of a group of covariates on the risk of disease, after adjusting for potential confounding factors, (see [23] Chapter 3 or [1] Chapter 10). When the binary response has $p$ predictors for the $i^{th}$ case-control or subject specific matched pair, we seek to model the logit of a grant of IP according to

$$logit(P(Outcome_{it} = 1) = X_{it}\boldsymbol{\beta} \qquad (1)$$

where $\boldsymbol{\beta}$ is a vector of $p$ parameters, $X_{it}$ is the associated set of model explanatory values and the occurrence is labelled $t = 0$ (control), $1$(case). When (1) is modelled via conditional maximum likelihood, [1] shows the resulting likelihood function is identical to one where the response (i.e. Outcome) is 1 for all matched pairs, while the predictors $(z_{p,i})$ are simply the difference between the case and control values of the explanatory variables (factor variables in Table 1 in our case), equivalently for the $p^{th}$ explanatory variable, $z_{p,i} = x_{p,i1} - x_{p,i0}$. Thus, we can generate conditional maximum likelihood estimates of $\boldsymbol{\beta}$ by running a logistic regression (e.g. using PROC LOGISTIC in SAS) on a response variable $y_i = 1$ (for all $i$ matched pairs), using explanatory variables $z_{p,i}$ that are simply the difference between the case and control measurements.

Crucially, the key focus of this study is to examine the latent effect of the Immigration Act 2004 as a policy instrument on IP outcomes. We operationalize this effect by introducing a factor variable 'After 2004' whose value is '1' after the year 2004 and '0' otherwise. Of course, it will be recalled we also used this variable as a matching variable when constructing our matched pairs. As this variable is identical for each case control pair, their difference is identically zero on all matched pairs. Accordingly, we cannot directly investigate the association between the Outcome and the Immigration Act. However, a clever way around this difficulty is suggested in [Sho], the idea is to investigate the possible interaction between each explanatory factor and the Immigration Act 2004. Thus, we create a set of interaction factors by forming product of each explanatory factor (called an original factor) with the 'After 2004' on each matched pair. By examining whether this product factor has a statistically significant effect of the Outcome, we can see whether the Immigration Act indirectly influences an explanatory factor's effect on the Outcome. Given we also matched on variables 'Nationality', 'Gender' and 'Adult', these are also excluded from the analysis by design. As a consequence the resulting logistic model incorporates 29 covariates, 16 of which are original factors listed in Table 1, while the remaining 13 are independent covariates that arise as interactions of these original factors with the 'After 2004' factor. We use PROC LOGISTIC in SAS to estimate the 29 parameters of this (conditional maximum likelihood) model. For completeness we provide descriptive statistics showing the proportion of cases having the attribute for each variable on the matched pairs dataset in Table 2. It is clear from this table that there is a considerable level of consistency between the cases and controls, reflecting the high degree of correlation induced by the propensity score matching mechanism.

| | Table 2 | | | |
|---|---|---|---|---|
| | Descriptive Statistics | | | |
| | Original Variable | | Original Variable by After 2004 Indicator | |
| Variable | Case | Control | Case | Control |
| SP Decision (Yes) | 0.01 | 0.02 | 0.01 | 0.02 |
| RIA (Yes) | 0.55 | 0.53 | 0.39 | 0.38 |
| Air Travel (Yes) | 0.43 | 0.42 | 0.15 | 0.14 |
| Asylum Reason (political) | 0.38 | 0.38 | 0.12 | 0.12 |
| Religion (stated) | 0.12 | 0.11 | 0.02 | 0.01 |
| Ethnicity (stated) | 0.29 | 0.30 | 0.04 | 0.03 |
| Unaccompanied Minor | 0.02 | 0.04 | 0.01 | 0.02 |
| Interviewed | 0.65 | 0.68 | 0.28 | 0.31 |
| Ever Married (Yes) | 0.50 | 0.49 | 0.22 | 0.22 |
| English Speaking (Yes) | 0.19 | 0.20 | 0.07 | 0.07 |
| Free Country of Origin | 0.41 | 0.42 | 0.15 | 0.16 |
| GDP | 0.19 | 0.21 | 0.06 | 0.06 |
| At Risk (political terror scale) | 0.67 | 0.67 | 0.32 | 0.32 |
| Refused Leave to Land | 0.09 | 0.10 | 0.09 | 0.10 |
| Returned to Country of Origin | 0.04 | 0.05 | 0.04 | 0.05 |
| Length (3 Years) | 0.05 | 0.06 | 0.03 | 0.04 |

The conditional maximum likelihood approach to estimating model (1) has the key feature that the results are conditional on the subject or person. Thus, within person correlation is accounted for and valid inferences for the model parameters at the population level are available. However, a drawback of model (1) is it does not allow us to condition on other key variables. As pointed out in section 2, nationality is central to the IP decision as it fixes the country of origin information. Accordingly, it makes sense to condition on it in our analysis. We therefore follow [19] and include nationality as a one-way random effect in model (1), giving a mixed model

$$logit(P(Outcome_{ijt} = 1)) = X_{it}\boldsymbol{\beta} + u_{ij} \qquad (2)$$

where $u_{ij} \sim N(0, \sigma_u^2)$, is the random effect for the subject, that is the person by nationality combination. We use PROC GLIMMIX in SAS to estimate the parameters (original factor and 'After 2004' interactions) of this mixed model.

## 5. Results and Analysis

Table 3 shows the results obtained from estimating model (1), the conditional maximum likelihood model, and Model (2), the mixed effects model, using the matched pairs dataset. Four columns are shown for each model. In each case the first column gives the parameter estimates $\widehat{\boldsymbol{\beta}}$ associated with each factor variable, while in the second the standard error of the parameter estimate is displayed. In the third column the significance level of the associated Wald test is given; a '*' is used to indicate significance at the $\alpha = 0.1\%$ level. Meanwhile, the final column for each model shows the odds ratio for each factor variable. For a dichotomous exploratory factor $x$ which takes the value 1 if the factor is present and 0 otherwise, the log of the odds ratio is given by

$$logit(P(Outcome = 1) - logit(P(Outcome = 0)) = (x = 1) \times \beta - (x = 0) \times \beta = \beta$$

Thus, the odds for a particular factor affecting the outcome are simply $\exp(\beta)$; it is this value that is reported in the fourth column.

Comparing the two sets of model results in Table 3 we can see considerable similarity but also some key differences. In particular the mixed model failed to produce meaningful estimates for five factors (shown with a '-');these factors are 'SP Decision', 'GDP', 'At Risk' and the latter pair crossed with the immigration act factor 'After 2004' respectively. Reasons for this failure include, the sub-population at subject level (i.e. person by nationality combination) for these factors is small, or there is linear dependence between original factors and their interaction with immigration act factor 'After 2004'. Meanwhile, for Model (1) the factors 'SP Decision' and 'Unaccompanied minor', and these two factors crossed with 'After 2004', have parameter estimates that are very large in absolute value with associated standard errors that are excessively large. So, if we exclude these seven factors, we find that both models produce a statistically significant result at the 5% level across 14 of the remaining 23 factors. Given this, it makes sense for us to draw our conclusion based on the set of 14 factors where both models agree. Furthermore, it is clear the estimates computed via model (1) are more variable than those from model (2). Interestingly, model (1) is a population average model across subjects (i.e. persons) while model (2) is subject specific at the person by nationality level. Since nationality is so vital to the determination process and the

Table 3
Model Fitting Statistics

| Variable | | Model (1) Logistic Model (Conditional Maximum Likelihood) | | | | Model (2) Logistic Mixed Model | | | |
|---|---|---|---|---|---|---|---|---|---|
| | | Parameter Estimate | Standard Error | Wald Test | Odds Ratio | Parameter Estimate | Standard Error | Wald Test | Odds Ratio |
| Original | SP Decision (Yes) | 30.1 | 332.2 | 0.93 | - | - | - | - | - |
| | RIA (Yes) | 1.19 | 0.09 | * | 3.28 | -0.20 | 0.04 | * | 0.82 |
| | Air Travel (Yes) | 0.04 | 0.03 | 0.20 | 1.04 | -0.22 | 0.03 | * | 0.80 |
| | Asylum Reason (political) | -0.04 | 0.03 | 0.23 | 0.96 | -0.30 | 0.03 | * | 0.74 |
| | Religion (stated) | 0.33 | 0.05 | * | 1.40 | 0.01 | 0.05 | 0.86 | 1.01 |
| | Ethnicity (stated) | -0.21 | 0.03 | * | 0.81 | -0.60 | 0.04 | * | 0.55 |
| | Unaccompanied Minor | -30.5 | 497.3 | 0.95 | - | -0.25 | 0.10 | 0.01 | 0.78 |
| | Interviewed | 1.43 | 0.10 | * | 4.16 | -0.25 | 0.03 | * | 0.78 |
| | Ever Married (Yes) | 1.08 | 0.08 | * | 2.95 | -0.11 | 0.03 | * | 0.89 |
| | English Speaking (Yes) | -0.27 | 0.05 | * | 0.76 | -0.32 | 0.04 | * | 0.72 |
| | Free Country of Origin | -0.02 | 0.12 | 0.86 | 0.98 | -0.30 | 0.03 | * | 0.74 |
| | GDP | -1.34 | 0.09 | * | 0.26 | - | - | - | - |
| | At Risk (political terror scale) | -0.31 | 0.09 | * | 0.74 | - | - | - | - |
| | Refused Leave to Land | -0.57 | 0.13 | * | 0.57 | -0.17 | 0.06 | 0.01 | 0.84 |
| | Returned to Country of Origin | -0.28 | 0.21 | 0.18 | 0.76 | 0.06 | 0.09 | 0.50 | 1.06 |
| | Length (Years) | 0.41 | 0.08 | * | 1.50 | 0.49 | 0.10 | 0.04 | 1.64 |
| Combination Factor: Original by 'After 2004' Indicator | SP Decision (Yes) | -14.5 | 224.8 | 0.95 | - | -0.24 | 0.11 | 0.03 | 0.79 |
| | RIA (Yes) | -1.52 | 0.12 | * | 0.22 | -0.39 | 0.05 | * | 0.68 |
| | Air Travel (Yes) | 0.01 | 0.05 | 0.77 | 1.01 | 0.27 | 0.05 | * | 1.30 |
| | Asylum Reason (political) | 0.12 | 0.05 | 0.02 | 1.12 | 0.31 | 0.06 | * | 1.36 |
| | Religion (stated) | 0.15 | 0.12 | 0.22 | 1.16 | 0.42 | 0.14 | * | 1.53 |
| | Ethnicity (stated) | 0.33 | 0.09 | * | 1.40 | 0.39 | 0.10 | * | 1.47 |
| | Unaccompanied Minor | 13.8 | 450.8 | 0.98 | - | -0.49 | 0.14 | * | 0.61 |
| | Interviewed | -2.60 | 0.13 | * | 0.07 | -0.22 | 0.05 | * | 0.80 |
| | Ever Married (Yes) | -1.42 | 0.10 | * | 0.24 | 0.08 | 0.05 | 0.15 | 1.08 |
| | English Speaking (Yes) | 0.25 | 0.08 | * | 1.28 | 0.25 | 0.07 | * | 1.29 |
| | Free Country of Origin | -0.61 | 0.19 | * | 0.54 | 0.13 | 0.05 | 0.01 | 1.14 |
| | GDP | 1.09 | 0.26 | * | 2.96 | - | - | - | - |
| | At Risk (political terror scale) | 0.05 | 0.14 | 0.71 | 1.05 | - | - | - | - |
| | Length (3 Years) | -0.96 | 0.10 | * | 0.38 | -1.13 | 0.13 | * | 0.32 |

* highlighted factors are statistically significant at the 0.1% level

mixed model explicitly takes this into account as a random effect, it is our view that model (2) is more appropriate for this matched pairs situation. Importantly, the contrast between the odds ratios across the 14 statistically significant factors for both models provides a useful insight into the impact of nationality on the IP outcome.

Looking at the results in upper part Table 3, we see there is an association between for the seven original statistically significant factors a grant of IP. These results are consistent with the analysis of this data as observational outcomes in [19]. However, as this is a matched pairs design we emphasise this is evidence for a contributory affect, rather than correlation as is the case in [19]. Accordingly, we can say that on average each of these seven factors in fact contributes to affecting a positive IP outcome. Under Model (1), conditional maximum likelihood based on the person only, population average odds ratios range from 0.57 to 4.16. Meanwhile, for mixed effects Model (2) the odds ratios range from 0.55 to 1.64 when the outcome is also conditioned on nationality. We note these estimates are sometimes referred to as shrinkage estimates, as their values are shrunk towards the mean (nationality) proportion by a shrinkage factor proportional to the size of each nationality sub-population. Indeed, shrinkage is often desirable as it protects against giving to much weight to small sub-populations. Clearly, the three largest odds ratios under Model (1) exceed 2.95, while the corresponding odds ratios for Model (2) are below 1.0. Thus, intriguingly, at a population level the factors 'RIA', 'Interviewed' and 'Ever Married' have a sizeable effect on the outcome; these findings are consistent with [19]. However, when we adjust the matched pairs for nationality, the effect disappears and being a RIA resident, or having had an interview or indeed being married may even reduce the odds of a grant of IP slightly. Seen in light of nationality, only the length of time in the protection process yields a substantial contributory effect of 1.64 on odds of a grant of IP. Ethnicity on the other hand yields odds of 0.55; this value is lower than the Model (1) value of 0.81 and lower than the value of 0.9 reported in [19]. Of course all these estimates are below 1, so importantly while their magnitude differs their direction is identical. On average therefore we can say that stating ethnicity contributes to reduce the chance of a grant of IP.

Of course the key focus of this study is whether the Immigration Act 2004 has a potential latent influence on IP outcomes. We operationalise this indirectly by examining the influence of each combination factor, listed in the bottom part of Table 3, on the odds of a grant of IP. Each combination factor is the product of an original factor with the factor 'After 2004', which determines whether the protection application was made after the introduction of the Immigration Act 2004. Table 3 shows that seven combination factors have a significant influence on the outcome. Once again we stress this is a matched pairs design so each combination factor contributes to either increasing or decreasing the odds of a grant of IP.

Specifically comparing the two sets of model results for the combination factors in Table 3, we can see there is once again considerable similarity across the odds ratios for both models. Under Model (1), conditional maximum likelihood, population average odds ratios range from 0.07 to 1.4, while for mixed effects Model (2) they range from 0.32 to 1.47. Once again it is clear the mixed model estimates are shrunk towards their mean nationality proportions. Examining the mixed model odds ratio estimates for the statistically significant combination factors, we can see a person stating their Ethnicity after the introduction of the Immigration Act 2004, has odds of a grant of IP that are almost 1.5 times those of a person who does not state their ethnicity or applied for protection prior to 2004. Thus in contrast to the analysis above, we find stating ethnicity proves beneficial to a grant outcome after the introduction of the Immigration Act 2004. Furthermore, the results show the combination factors 'Asylum Reason', 'English Speaking' and 'Free Country of Origin' also yield odds ratios greater than 1.0. Thus, for these four combination factors the introduction of the Immigration Act 2004 coincided with, and in fact contributed to, increasing the chance of a grant of IP. This is a significant finding. On the other hand the Immigration Act impacted adversely on those who stayed in a reception centre 'RIA', or had an interview or those waiting in the IP determination process for at least three years. Indeed the results show the odds ratio for the combination factor involving length of stay in the IP determination process is only 0.32. Thus, while the effect of original length of stay variable considered above had a positive influence on the IP outcome, we see the introduction of the Immigration Act 2004, contributed to a situation where the chances of a positive IP outcome are about three times poorer on average.

Clearly, these findings relating to the impact of the Immigration Act 2004 are both novel and profound. If the introduction of the Immigration Act 2004 had increased the odds of a grant of IP across all seven factors, then we could conclude the Immigration Act 2004 has worked in favour of those seeking protection. This in turn would have caused the recognition rate to increase for those seeking protection. Equally, had the odds of a grant of IP fallen across all seven factors, then we might conclude the Immigration Act 2004 has had an adverse influence on those seeking protection. However, neither of these situations occurs and on balance the Immigration Act has had both a positive and negative effect on IP outcomes. Accordingly, while in [18] and [19] there is evidence for a stiffer protection regime in Ireland in later years, this analysis shows there is little justification for the view that this is caused by immigration policy. Based on this it seems unlikely that immigration policy is a latent tool affecting the protection process in Ireland. Moreover, this implies that human rights principles embodied in the GC do not appear to be infringed upon by state actors managing asylum and immigration processes in Ireland. Indeed it would suggest that state actors are mindful of complying with the spirit of the GC while having due regard to the rights of citizens.

In section 2 we also highlighted the importance of overall average odds associated with statistically significant factors considered as a group. In particular we drew attention to the overall average effect of factors associated with the IP procedure relative to average effect of applicant specific factors. We noted if this ratio was large then immigration is likely being used to limit access to protection. In this instance the statistically significant factors associated with the person are 'Asylum Reason', 'Ethnicity' and 'English Speaking', while those associated with the procedure are 'Interviewed' and 'Free Country of Origin'. The average odds for both groups are 1.32 and 0.92 respectively, so their ratio is 1.44. Thus, overall the odds that applicant factors determine the outcome are almost one-and-a-half times those of procedural factors. This shows that while immigration acts to influence both applicant specific and procedural factors, the overall odds of a grant of IP are lower for the latter. Accordingly, this weighting in favour of the applicant suggests that immigration is not used as a latent tool for managing access to the asylum process. Once again this points to the likelihood that officials view protection applicants humanely and therefore tend to comply with and are committed to the spirit of IP norms and procedures.

In general the findings of this analysis reinforce some of the key findings of other studies such as [2], [3], [14] and [19]. Broadly speaking there is agreement that nationality is the primary axis upon which a protection determination revolves, and when it is adjusted for determination outcomes are for the most part fair. As noted in [19] this fact serves to strengthen our faith in protection procedures in Ireland. Accordingly, it validates the efforts of those officials charged with managing, monitoring and coordinating protection processes at national, EU and international levels. Moreover, as stressed here, this is analysis is based on a reconfiguration of observational data into matched pairs. When viewed in this light it is clear that nationality in fact is the key factor contributing to the odds of a grant of IP. Indeed, if it was otherwise the IP determination procedure in Ireland would be profoundly flawed. The fact that nationality is so central in this matched pairs set up assures the IP determination procedure in Ireland is credible. In addition, above we observed greater variability in the odds ratios for Model (1), 0.07 to 1.4, as compared to Model (2), 0.32 to 1.47. This indicates the shrinkage estimates for the odds of a grant of IP are fairly smooth functions when we adjust for nationality. In light of the comments at the end of section 2, this implies the odds for an IP grant for specific nationalities are not adversely influenced by latent immigration policies. This adds further weight to the overall picture arising from this analysis that IP is not influenced by immigration control actions.

## 6. Conclusions

In this study we have examined over 40,000 individual IP outcomes for non-EEA nationals covering a 16 year period in Ireland with a view to seeing whether immigration policy may have impacted on Ireland's generosity to asylum seekers. In particular we have considered this question in the realm of the international relations debate that surrounds protection. In order to address the issue we have reconfigured the individual outcomes into a set of over 23,000 matched pairs based on direct matching by nationality, gender, whether an adult and whether the application was made after the introduction of the Immigration Act 2004. For the remaining variables we used the propensity score matching approach set out in [10]. A key feature of this approach is that it replicates the statistical features of an experimental set-up where observational data only are to hand. Importantly, the analysis we conduct based on match pairs allows to say the explanatory factors are not just correlated but in fact contribute to the grant of IP. We emphasise this is a key innovation in the analysis of protection outcomes that, to our knowledge, has not been conducted elsewhere.

We modelled our matched pairs in two ways, via conditional maximum likelihood and using a mixed model. Our view is the mixed model is a somewhat better approach as it allows us to account for person and nationality specific effects. Nonetheless, we found that both models gave similar results in terms of the factors that are statistically significant. More interestingly, the analysis showed that when we adjusted for nationality only the length of time in the IP process contributed to improving the odds of an IP outcome.

Of course the main focus of this study was to see if there was evidence that immigration policy is a latent tool used to influence protection application numbers. Specifically, did the introduction of the Immigration Act 2004 influence the odds of a grant of protection? On balance we found this was not the case, a conclusion that is both novel and profound in a matched pair context. Accordingly, there can be little justification for the view that immigration policy is a latent tool affecting the protection process in Ireland even though the IP regime in Ireland is stiffer in later years. This indeed implies that human rights principles do not appear to be infringed upon by state actors in the immigration sphere in Ireland. In fact it seems more likely that state actors try to comply with the spirit of the GC. Constructivists and realists can take comfort from this inference since if it were otherwise the determination system in Ireland would be deliberately biased. Furthermore, this view of the IP regime in Ireland also suggests there is little justification for a belief in the notion of "fortress Europe", since if this was the case, then odds for a grant of IP could only be adversely influenced by immigration control measures. On this basis it also seems that concerns about burden sharing do not play on the minds of officials making IP determinations in Ireland.

We also examined the influence of a group of applicant specific factors relative to procedural factors on the odds of a grant of IP. We found the overall the odds that applicant factors determine the outcome are almost one-and-a-half times those of procedural factors. On this basis we suggested that immigration policy does not influence protection decisions and this pointed to the interpretation that officials tended to be open-minded toward protection applicants and viewed there application humanely. Further, when we examined the influence of nationality with the context of the matched pairs configuration, we found it is the main contributory element in a determination, this was expected and attests to the credibility of the protection system in Ireland. In this specific context realist state actors are therefore making determinations in accordance with the international norms. Indeed, based on the shrinkage estimates for the odds of a grant of IP from Model (2) we felt that specific nationalities are not adversely influenced by latent immigration policies. Based on these findings, decisions of the Irish protection system are credible and do not play out like a system of 'refugee roulette' as described in [22].

On a more general, level asylum seekers instinctively feel themselves to be in refugee like situations because the humanitarian rights in their home country often fall short of those of citizens of liberal western democracies, who by comparison are free politically, economically and socially. On this basis many would be asylum seekers, who may not be refugees in the strict sense, might hope that most western democracies will treat them in a humanitarian way and in the spirit of the GC. When determination officers are face-to-face with asylum seekers our analysis supports this conclusion. Moreover, our key conclusion that latent immigration policy does not influence protection decisions, indicates that on the whole Ireland is 'open' to the concerns of those in need of protection and therefore does not use immigration control to dodge the thorny issue of burden-sharing. Meanwhile, [18] and [19] showed the determination procedure in Ireland is stiffer in later years. However here using matched pairs there is little contributory evidence to support this view in the context in immigration control. This too would suggests Ireland is acting according to the spirit of the GC and living up to it international commitments.